\documentclass[10pt]{article} 
\usepackage{comb, amsmath, latexsym, amssymb, amscd}
\numberwithin{equation}{section}

\begin{document}

\title{An invitation to quantum tomography}
\author{ Richard Gill
\footnote{Mathematical Institute, University of Utrecht, 
Box 80010, 3508 TA Utrecht, The Netherlands, gill@math.uu.nl, http://www.math.uu.nl/people/gill}
\footnote{Eurandom, P.O. Box 513, 5600 MB Eindhoven, The Netherlands, 
guta@eurandom.tue.nl, http://euridice.tue.nl/~mguta/} 
~ and M\u{a}d\u{a}lin Ionu\c{t} Gu\c{t}\u{a}$^\dag$}
\date{}
\maketitle


\vspace{8mm}

\begin{abstract} 
We describe quantum tomography as an inverse statistical problem
and show how entropy methods can be used to study the behaviour of
sieved maximum likelihood estimators. There remain many open problems,
and a main purpose of the paper is to bring these to the attention of 
the statistical community.
\end{abstract}

\section{Introduction}
\label{sec.introduction}

It is curious that it took more than eighty years from its discovery
till it was possible to experimentally determine and visualize the most
fundamental object in quantum mechanics, the wave function. The forward
route from quantum state to probability distribution of measurement
results has been the basic stuff of quantum mechanics textbooks for
decennia. That the corresponding mathematical inverse problem had a
solution, provided (speaking metaphorically) that the quantum state has
been probed from a sufficiently rich set of directions, had also been
known for many years. However it was only in 1993, with  \cite{Smithey},
that it became
feasible to actually carry out the corresponding measurements on one
particular quantum system---in that case, the state of one mode of
electromagnetic radiation (a pulse of laser light at a given frequency).
The resulting pictures have since made it to the front covers of
journals like Nature and Science, and experimentalists use the technique
to establish that they have succeeded in creating non-classical forms of
laser light such as squeezed light and Schr\"odinger cats. The
experimental technique we are referring to here is called quantum
homodyne tomography: the word homodyne referring to a comparison between
the light being measured with a reference light beam at the same
frequency. We will explain the word tomography in a moment.

The quantum state can be represented mathematically in many different
but equivalent ways, all of them linear transformations on one another.
One favourite is as the Wigner function $W$: a real function of two
variables, integrating to plus one over the whole plane, but not
necessarily nonnegative. It can be thought of as a ``generalised joint
probability density'' of the electric and magnetic fields, $q$ and $p$.
However one cannot measure both fields at the same time and in quantum
mechanics it makes no sense to talk about the values of both electric
and magnetic fields simultaneously. It does, however, make sense to talk
about the value of any linear combination of the two fields, say $\cos
(\phi) q + \sin (\phi) p$ (in fact, one can think of $\phi$ as simply representing 
time). And one way to think about the statisticial
problem is as follows: the unknown parameter is a joint probability
density $W$ of two variables $Q$ and $P$. The data consists of independent
samples from the distribution of $(X,\Phi)=(\cos (\Phi) Q + \sin
(\Phi) P,\Phi)$, where $\Phi$ is chosen independently of $(Q,P)$, and uniformly in
the interval $[0,\pi]$. Write down the mathematical model expressing the
joint density of $(X,\Phi)$ in terms of that of $(Q,P)$. Now just allow
that latter joint density, $W$, to take negative as well as positive values
(subject to certain restrictions which we will mention later). And that
is the statistical problem of this paper.

This is indeed a classical tomography problem: we take observations from
all possible one-dimensional projections of a two-dimensional
probability density. The non-classical feature is that though all these
one-dimensional projections are indeed bona-fide probability densities,
the underlying two-dimensional ``joint density'' need not itself be a
bona-fide joint density, but can have small patches of ``negative
probability density''.

Though the parameter to be estimated may look weird from some points of
view (for instance, when one looks at it ``as a probability density''),
it is mathematically very nice from other points of view. For instance,
one can also represent it by a matrix of (a kind of) Fourier
coefficients: one speaks then of the ``density matrix'' $\rho$. This is an
infinite dimensional matrix of complex numbers, but it is a positive and
selfadjoint matrix with trace one. The diagonal elements are real
numbers summing to one, and forming the probability distribution of the
number of photons found in the light beam (if one could do that
measurement). Conversely, any such matrix $\rho$ corresponds to a physically
possible Wigner function $W$, so we have here a concise 
mathematical characterization of
precisely which ``generalized joint probability densities'' can occur.

The initial reconstructions were done by borrowing analytic techniques
from classical tomography---the data was binned and smoothed, the
inverse Radon transform carried out, followed by some Fourier
transformations. At each of a number of steps, there are numerical
discretization and truncation errors. The histogram of the data will
not lie in the range of the forward transformation (from quantum state
to density of the data). Thus the result of blindly applying an inverse
will not be a bona-fide Wigner function or density matrix. Moreover the
various numerical approximations all involve arbitrary choices of
smoothing, binning or truncation parameters. Consequently the final
picture can look just how the experimenter would like it to look and
there is no way to statistically evaluate the reliability of the result.
On the other hand the various numerical approximations tend to destroy
the interesting ``quantum'' features the experimenter is looking for, so
this method lost in popularity after the initial enthousiasm.

So far there has been almost no attention paid to this problem by
statisticians, which is a shame, since on the one hand it is one of the
most important statistical problems coming up in modern physics, and on
the other hand it is ``just'' a classical nonparametric statistical
inverse problem. The unknown parameter is some object $\rho$,
or if you prefer $W$, lying in an
infinite dimensional linear space (the space of density matrices, or the
space of Wigner functions; these
are just two concrete representations in which the experimenter has
particular interest). The data has a probability distribution which
is a linear transform of the parameter. Considered as an analytical
problem, we have an ill-posed inverse problem, but one which has a lot
of beautiful mathematical structure and about which a lot is known (for
instance, close connection to the Radon transform). Moreover it has
features in common with nonparametric missing data problems (the
projections from bivariate to univariate, for instance, and there are
more connections we will mention later) and with nonparametric density
and regression estimation. Thus we think that the time is ripe for this
problem to be ``cracked'' by mathematical and computational
statisticians. In this paper we will present some first steps in that
direction.

Our main results will be consistency theorems for two estimators. Both
estimators are based on approximating the infinite dimensional parameter
$\rho$ by a finite dimensional parameter, in fact, thinking of $\rho$ as
an infinite dimensional matrix, we simply truncate it to an $N\times N$
matrix where the truncation level $N$ will be allowed to grow with the
number of observations $n$. The first estimator employs some analytical
inverse formulas expressing the elements of $\rho$ as mean values of
certain functions of the observations $(X,\Phi)$. Simply replace the theoretical
means by empirical averages and one has unbiased estimators of the
elements of $\rho$, with moreover finite variance. If one applies this
technique without truncation the estimate of the matrix $\rho$ as a
whole will typically not satisfy the nonnegativity constraints. The
resulting estimator will not be consistent either, with respect to
natural distance measures. But provided the truncation level grows with
$n$ slowly enough, the truncated estimate will satisfy the constraints,
and provided it grows fast enough, the overal estimator will be
consistent.

There are many unbiased estimators of the matrix elements of $\rho$ and
the choice we make is based on analytic tractability, not on any
optimality criteria.

The second estimator we study further exploits the same idea, in a more
canonical way: we study the sieved maximum likelihood estimator based on
the same truncation to a finite dimensional problem. The truncation
level $N$ needs to depend on sample size $n$ to balance bias and
variance. We prove consistency of the sieved mle under an appropriate
choice of $N(n)$ by applying a general theorem of \cite{Wong&Shen}. In
order to verify the conditions we need to bound certain metric entropy
integrals (with bracketing) which express the size
(infinite-dimensional-ness) of the statistical model under study.

This turns out to be feasible, and indeed to have an elegant solution,
by exploiting features of the mapping from parameters (density matrices)
to distributions of the data. Various distances between probability
distributions possess analogues as distances between density matrices,
the mapping from parameter to data turns out to be a contraction, so we
can bound metric entropies for the statistical model for the data with
quantum metric entropies for the class of density matrices. And the
latter can be calculated quite conveniently.

Our results form just a first attempt at studying the statistical
properties of estimators which are already being used by experimental
physicists, but they show that the basic problem is both rich in
interesting features and tractable to analysis. The main result so far
is a consistency theorem for a sieved maximum likelihood estimator,
which depends on an assumption of the rate at which a truncated density
matrix approximates the true one. It seems that the assumption is
satisfied for the kinds of states which are met with in practice.
However, further work is needed here to describe in physically
interpretable terms, when the estimator works. Secondly, we need to
obtain rates of consistency and to further optimize the construction of
the estimator. Thirdly, one should explore the properties of penalized
maximum likelihood, and if possible make it adaptive to the rate of
approximation of the truncated model, so that the truncation level
$N(n)$ is determined from the data.

We largely restrict attention to an ideal case of the problem where
there is no further noise in  the measurements. In practice, the
observations have added to them Gaussian disturbances of known variance.
There are some indications that when the variance is larger than a
threshold of $1/2$, reconstruction becomes impossible or at least,
qualitatively much more difficult. This needs to be researched from the
point of view of optimal rates of convergence. The threshold should not
be an absolute barrier for sieved or penalized maximum likelihood, though 
it may well have qualitative impact on how well this works.

We also only considered one particular though quite convenient way of
sieving the model, i.e., one particular class of finite dimensional
approximations. There are many other possibilities and some of them
might allow easier analysis and easier computation. For instance,
instead of truncating the matrix $\rho$ in a given basis, one could
truncate in an arbitrary basis, so that the finite dimensional
approximations would corespond to specifying $N$ arbitrary state vectors
(eigenvectors) and a probability distribution over these ``pure
states''. Now the problem has become a missing data problem, where the
``full data'' would assign to each observation also the label of the
pure state from which it came. In the full data problem we need to
reconstruct not a matrix but a set of vectors, together with an ordinary
probability distribution over the set, so the ``full data'' problem is statisticially
speaking a much easier problem that the missing data problem. One could
imagine that the EM algorithm, or Bayesian reconstruction methods, could
exploit this structure.

We concentrated on estimation of $\rho$ but it would also be interesting to 
obtain results on estimation of $W$. The analogy with density estimation could
suggest new statistical approaches here.
Finally, it is most important to add to the estimated parameter, estimates of its
accuracy. This is absolutely vital for applications, but so far no valid approach 
is available.

The quantum mathematical physics of this problem is identical to that of
the quantum simple harmonic oscillator, where $q$ and $p$ stand for 
position and momentum of a particle, oscillating inside a quadratic potential well.
In the next section we describe this mathematics using the terminology of
position and momentum.

Section 3 is devoted to the ad hoc estimator based on truncation of $\rho$, and 
Section 4 to the sieved maximum likelihood estimator. That section finishes with some
concluding remarks to the whole paper.

\nopagebreak

\section{Quantum systems and measurements}
\label{sec.qmeasurements}

In classical mechanics the state of macroscopic systems like billiard balls, 
pendulums or stellar systems is described by its ``coordinates'' in a phase space, each 
coordinate corresponding to an attribute which we can measure such as position and momentum. 
Therefore the functions on the phase space are called observables. 
When there exists uncertainty about the exact point in the phase space, or we deal with a 
statistical ensemble, the state is a probability distribution and the observables become random 
variables. 
Quantum mechanics also deals with observables such as position and momentum of a particle, 
spin of an electron, number of photons in a cavity but 
breaks from classical mechanics in that these are no longer functions but {\it selfadjoint 
operators} on a complex Hilbert space $\mathcal{H}$ with inner product 
$\br\cdot,\cdot\ke$ which is linear in the right slot and anti-linear in the left one. 
For example, the components in different 
directions of the spin of an electron are certain selfadjoint operators on $\C^2$, 
or hermitian $2\times 2$ matrices which do not commute with each other. 
Another quantum system with which we will deal in this paper is the 
quantum particle. Its basic observables position and momentum, are two unbounded 
selfadjoint operators 
$\mathbf{Q}$ and $\mathbf{P}$ respectively acting on the complex Hilbert space 
$\mathcal{H}=L^2(\R)$ as 
\begin{eqnarray} 
&&(\mathbf{Q}\psi_1)(x) = x\psi_1(x),\nonumber\\
&&(\mathbf{P}\psi_2)(x) = -i\frac{d\psi_2(x)}{dx},\nonumber
\end{eqnarray}
for $\psi_1,\psi_2$, vectors in their respective domains. The operators satisfy Heisenberg's 
{\it canonical commutation relations} $\mathbf{Q}\mathbf{P}-\mathbf{P}\mathbf{Q}=i\mathbf{1}$. 
We note that the algebra generated by all bounded functions of $\mathbf{Q}$ and 
$\mathbf{P}$ is dense in  the space of bounded operators $\mathcal{B}(\mathcal{H})$ 
with respect to the weak operator topology, defined by the seminorms 
$|\br \psi, \cdot\psi\ke|$ for all $\psi\in\mathcal{H}$. 
For this reason $\mathcal{B}(\mathcal{H})$ is 
usually considered the algebra of observables of the system.

The state of the quantum system is given by a {\it density matrix} $\rho$, i.e. a 
positive trace-class operator with $\Tr (\rho)=1$. This is analogue to the probability 
distribution on the phase space, the expectation of an observable 
$\mathbf{X}\in\mathcal{B}(\mathcal{H})$ being given by 
$\mathbb{E}_\rho(\mathbf{X}):=\mathrm{Tr}(\rho \mathbf{X})$. 
The states form a convex subset $\mathcal{S}(\mathcal{H})$ of the space of trace-class 
operators on $\mathcal{H}$, the latter being denoted $\mathcal{T}_1(\mathcal{H})$. The 
lower script $1$ refers to the norm 
\begin{equation}
\|\tau\|_1:=\mathrm{Tr}(|\tau|),
\end{equation} 
with respect to which $\mathcal{T}_1(\mathcal{H})$ is a Banach space. 
If $\tau$ is selfadjoint then it can be represented as an infinite diagonal matrix 
(in a certain basis of the space $\mathcal{H}$) with elements $\tau_i$, thus 
$\|\tau\|_1:\sum_i|\tau_i|$. Any state can be written, in general non-uniquely, 
as convex combination of {\it pure} or vector states which have expectations of the form 
\begin{equation}
\mathbb{E}_\psi(\mathbf{X}):=\mathrm{Tr}(\mathbf{P}_\psi \mathbf{X})=\br\psi,\mathbf{X}\psi\ke 
\end{equation}
where $\mathbf{P}_\psi$ is the orthogonal projection on the space $\C\psi$. 
There exists a duality relation 
\begin{equation*}
\mathcal{T}_1(\mathcal{H})^*=\mathcal{B}(\mathcal{H})
\end{equation*}
which is the non-commutative analogue of $\ell_1^*=\ell_\infty.$

But how do we actually measure an observable of the system? This is in general a difficult 
question from the practical point of view, as we will see in this paper only certain observables 
can be measured with the present technology. But we can describe how the probability distribution 
of the results will look if we perform the measurement. Any selfadjoint operators $\mathbf{X}$ 
has a spectral decomposition or ``diagonalization'' 
\begin{equation*}
\mathbf{X}= \sum_{i\in\sigma(\mathbf{X})}x_i \mathbf{P}_i
\end{equation*}
where the sum is taken over the spectrum of $\mathbf{X}$ and $\mathbf{P}_i$ is the projection 
associated to the eigenvalue $x_i$. The sum should be replaced by an integral for operators with 
continuous spectrum. If the system is in the state $\rho$ then the probability of obtaining 
the value $x_i$ is 
\begin{equation}
p_\rho(i)=\Tr(\rho \mathbf{P}_i).
\end{equation} 
which depends only on the spectral projections, the eigenvalues $x_i$ being just labels of the 
results. More realistic measurements are modeled by 
{\it positive operator valued measures} (POVM) which are maps $\mathbf{M}$ from the 
$\sigma$-algebra of a measure space $(\Omega_\mathbf{M},\Sigma_\mathbf{M})$ into 
$\mathcal{B}(\mathcal{H})$ with the following properties: $\mathbf{M}(A)=\mathbf{M}(A)^*\geq 0$ 
for any $A\in\Sigma_\mathbf{M}$, $\mathbf{M}(\cup_i A_i)=\sum_i \mathbf{M}(A_i)$ 
for a countable number of arbitrary disjoint $A_i\in\Sigma_\mathbf{M}$, 
and $\mathbf{M}(\Omega_\mathbf{M})=\mathbf{1}$. 
Similarly to the projection valued case, the probability distribution of the results is 
\begin{equation*}
P_{\rho}^{(\mathbf{M})}(A)=\mathrm{Tr}(\rho \mathbf{M}(A)).
\end{equation*}
An important feature of the map $\rho\mapsto P_{\rho}^{(\mathbf{M})}$ is that it is 
{\it contractive} in appropriate norms. The total variation distance between two probability 
distributions on $(\Omega_\mathbf{M},\sigma_\mathbf{M})$ is defined by  
\begin{equation}\label{eq.total.variation.distance}
d_{\text{tv}}(P_1,P_2):=\sup_{|F|\leq 1}\left|\int F(x)P_1(dx)-\int F(x) P_2(dx)\right|
\end{equation} 
Then 
\begin{eqnarray} \label{eq.contactivity.measurements}
&&d_{\text{tv}}\left(P_{\rho}^{(\mathbf{M})},P_{\rho'}^{(\mathbf{M})}\right)
= \sup_{|F|\leq 1}
\left|\int F(x)P_{\rho}^{(\mathbf{M})}(dx)-\int F(x)P_{\rho'}^{(\mathbf{M})}(dx)\right|\nonumber\\ 
&&= \sup_{|F|\leq 1}\left|\Tr((\rho-\rho')\int F(x)\mathbf{M}(dx))\right|
\leq \|\rho-\rho'\|_1\nonumber ,
\end{eqnarray}
where in the last step we have used the fact that $\int F(x)\mathbf{M}(dx)\leq\mathbf{1}$ and 
then we applied the inequality $|\Tr(\tau \mathbf{Y})|\leq\|\tau\|_1\|\mathbf{Y}\|$ for all $\tau$ 
trace-class and $\mathbf{Y}$ bounded, in which the reader recognizes its classical counterpart 
$|\int fg |\leq \|f\|_1\|g\|_\infty$.

Notice that we are merely concerned here with the distribution of the results and 
do not specify the state of the system after the measurement. The ``no quantum cloning theorem'' 
shows that measurements on a single system cannot completely reveal its state, in other words 
if the state is left unchanged after measurement then the results do not give us any 
information on the state. 

\vspace{4mm}

We can now formulate our problem in the following way: we have at our disposal a large number of 
systems identically prepared in an unknown state $\rho$, on each one of them we can perform a 
certain measurement, and we want to construct an estimator of $\rho$ based on the 
measurement results. Suppose for simplicity that we make the same measurement $\mathbf{M}$ 
on all particles, then the results are i.i.d. random variables $X_1,X_2,\dots$ on 
$(\Omega_\mathbf{M},\Sigma_\mathbf{M})$ with distribution $p_{\rho}^\mathbf{M}$. We will be 
interested in identifiable models, meaning that the map 
$\mathbf{T}_\mathbf{M}:\rho\mapsto p_{\rho}^\mathbf{M}$ is one-to-one. 
For further details on quantum statistical inference we refer to the review 
\cite{Barndorff-Nielsen&Gill&Jupp} and the classical textbook \cite{Holevo}.

\section{Quantum homodyne tomography}
\label{sec.qhomodyne}

Let us return to the quantum particle described by the observables $\mathbf{Q}$ and $\mathbf{P}$ 
satisfying the canonical commutation relations 
$\mathbf{Q}\mathbf{P}-\mathbf{P}\mathbf{Q}=\mathbf{1}$. The problem of measuring observables 
other than position and momentum has been elusive until ten years ago when pioneering 
experiments in quantum optics conducted by Raymer's group \cite{Smithey} lead to a powerful 
measurement technique called {\it homodyne tomography}. The quantum system to be measured 
is laser light with a fixed frequency whose observables are the field amplitudes satisfying 
commutation relation identical to those which characterize the quantum particle. 
Their linear combinations 
$\mathbf{X}_\phi=\cos \phi\mathbf{Q}+\sin\phi \mathbf{P}$ are called {\it quadratures}, and 
homodyne tomography is about measuring the quadratures for an {\it arbitrary} 
phase $\phi\in [0,\pi]$. 
The experimental setup consists of an additional laser of high intensity called local oscillator 
(LO), which is combined with the mode of unknown state through a fifty-fifty beam splitter, 
and two photon detectors each one measuring one of the emerging beams. 
Then a rescaled difference of the measurement results turns out to have the same probability 
distribution as that of the quadrature $\mathbf{X}_\phi$ in the limit of infinite intensity LO. 
It can be shown that the probability distribution $P_\rho(\cdot,\phi)$ on $\R$ has density 
$p_\rho(\cdot,\phi)$ with respect to the 
Lebesgue measure and generating function  
\begin{equation}\label{eq.generating_function}
\mathbb{E}(e^{itX}|\phi)=\mathrm{Tr}(\rho e^{it\mathbf{X}_\phi}). 
\end{equation}
The phase $\phi$ is controlled by the experimenter by adjusting a parameter of the local 
oscillator, and it will be assumed to be chosen randomly uniformly distributed over the 
interval $[0,\pi]$. Then the joint probability distribution for the pair consisting in 
measurement results and phases $(X,\Phi)$ has density $p_\rho(x,\phi)$ 
with respect to the measure $dx\times\frac{d\phi}{\pi} $ on $\R\times[0,\pi]$. 
A natural way of representing the state $\rho$ is by writing down its matrix elements 
$\rho_{i,j}:=\br \psi_i,\rho\psi_j\ke$ in an orthonormal basis of the Hilbert space $L_2(\R)$, 
for example
\begin{equation}\label{eq.psi_n}
\psi_i(x):=\frac{H_i(x)}{(\sqrt{\pi}~2^i i!)^{1/2}}~e^{-x^2/2},
\end{equation}
where $H_i$ are Hermite polynomials. This basis has a special relevance in quantum optics, 
$\psi_i(x)$ being pure states of exactly $i$ photons. 
Here is the concrete formula for $p_\rho$ in terms of $\rho_{j,k}$:
\begin{equation} \label{eq.concrete.p}
p_\rho(x,\phi)=\sum_{j,k=0}^\infty \rho_{j,k}\psi_k(x)\psi_j(x)e^{-i(j-k)\phi}.
\end{equation}

An important feature of this homodyne detection scheme is the invertibility of the map 
$\mathbf{T}:\rho\to p_\rho(\cdot,\cdot)$, making it theoretically possible to 
infer the state of the system from the knowledge of the distribution of results. This was not 
possible had we measured only a finite number of quadratures! But what is the connection 
of this method with the more familiar computerized tomography used in the hospitals? 
Well, physicists like to  represent the state of a quantum system by a certain function on $\R^2$ 
called the {\it Wigner function} $W(q,p)$ which is much like a joint probability distribution for 
$\mathbf{P}$ and $\mathbf{Q}$ in the sense that its marginals are the probability distributions 
for measuring $\mathbf{Q}$ and respectively $\mathbf{P}$. Of course the two observables cannot 
be measured simultaneously so we cannot speak of a joint distribution, in fact the Wigner 
function need not be positive but many interesting features of the quantum state can be 
visualized in this way. It turns out that $p_\rho(x,\phi)$ is the {\it Radon transformation} 
of the Wigner function
\begin{equation*}
p_\rho(q,\phi)=\int_{-\infty}^\infty W(q\cos\phi+p\sin\phi, q\sin\phi-p\cos\phi)dp.
\end{equation*}
The Radon transformation and its inverse play a distinguished role in computerized tomography. 
Here one reconstructs a ``shape'', for example the spatial distribution of the 
absorption coefficient for X-ray in a cross-section of the human body, 
by recording the transmitted radiation along an axis perpendicular to the beam and repeating this 
with the apparatus rotated at different angles. In our case the Wigner function is the unknown 
function while the probability density $p_\rho$ represents the transmitted angle-dependent 
signal. The term optical homodyne tomography was coined in 1993 \cite{Smithey} when the 
first Wigner function was reconstructed from experimental data using the homodyne scheme. 
The Fourier transform of the Wigner function has the following expression 
\begin{equation*}
\widetilde{W}(u,v)=\mathrm{Tr}\left(\rho e^{-iu\mathbf{Q}-iv\mathbf{P}}\right),
\end{equation*}
and we note that if $\mathbf{Q}$ and $\mathbf{P}$ were commuting operators then 
$W(q,p)$ would indeed be the joint probability distribution of outcomes of their measurement. 
Finally, from 
$\tilde{W}(u,v)$ we can obtain the matrix elements of the state $\rho$ with respect to 
a fixed orthonormal basis by integrating with certain kernel functions \cite{Leonhardt}. 
In practice this procedure has its drawbacks 
because it involves ``filtering'' the data as in usual tomography which as argued in 
\cite{D'Ariano.0} amounts to tampering with the state that is, making it more ``classical''.  

\vspace{4mm}  

In \cite{D'Ariano.0} D'Ariano et al. presented a technique which provides the 
matrix elements without calculating the Wigner function as an intermediary step. 
The method has been further analyzed in \cite{D'Ariano.1, D'Ariano.2, D'Ariano.3}. 
The key formula shows that any operator $\tau\in\mathcal{T}_1(\mathcal{H})$ can be expressed 
as a linear superposition of functions of the observables $\mathbf{X}_\phi$: 
\begin{equation}\label{eq.quantum_tomographic_rho}
\tau=\frac{1}{4}\int_{-\infty}^{\infty} dr |r| \int_0^\pi \frac{d\phi}{\pi}
\Tr (\tau e^{ir\mathbf{X}_\phi}) e^{-ir\mathbf{X}_\phi}.
\end{equation}
which is an application of the general 
theory of {\it quantum tomography} developed 
by D'Ariano and his collaborators \cite{D'Ariano.8, D'Ariano.10}. 
By applying this formula to the state $\rho$ and using (\ref{eq.generating_function}) we get 
\begin{equation*}
\rho=\int_{-\infty}^{\infty}dx \int_0^\pi \frac{d\phi}{\pi}
p_\rho(x,\phi)K(x-\mathbf{X}_\phi),
\end{equation*}
where $K$ is the generalized function given by 
\begin{equation}\label{eq.ker1}
K(x)=-\frac{1}{2}\mathcal{P}\frac{1}{x^2}=
-\lim_{\epsilon\to 0^+}\mathrm{Re}\frac{1}{(x+i\epsilon)^2}.
\end{equation}
In order to obtain a mathematically sound expression we take the matrix elements on both side 
\begin{equation}\label{eq.quantum_tomographic_rho_n,n+d}
\rho_{k,j}=\int_{-\infty}^\infty dx \int_0^\pi \frac{d\phi}{\pi} p_\rho(x,\phi)
f_{k,j}(x)e^{-i(j-k)\phi}, 
\end{equation}
with $f_{k,j}$ bounded functions which in the quantum tomography literature are called 
{\it pattern functions}. A first concrete expression using Laguerre polynomials 
was found in \cite{D'Ariano.5}, and was followed by a more transparent one \cite{Leonhardt.Richter}
\begin{equation}\label{eq.pattern_functions_leonhardt}
f_{k,k+d}(x,\phi)=e^{-id\phi}\frac{d}{dx}(\psi_k(x)\varphi_{d+k}(x)),
\end{equation}
in terms of the basis vectors $\psi_k$ and a certain un-normalizable solution of the 
Scr\"{o}dinger equation
\begin{equation}
\left[-\frac{1}{2}\frac{d^2}{dx^2}+\frac{1}{2}x^2\right]\varphi_{j}=\omega_j~\varphi_j,
\end{equation}.

Equation (\ref{eq.quantum_tomographic_rho_n,n+d}) suggests the {\it unbiased estimator} 
of $\rho$ based on the first $n$ i.i.d. results $(X_l,\Phi_l)$ 
whose matrix elements are 
\cite{D'Ariano.2, D'Ariano.3, Leonhardt.Richter}:
\begin{equation}\label{eq.unbiased_estimator}
\hat{\rho}^{(n)}_{k,j}=\frac{1}{n}\sum_{l=1}^n f_{k,j}(X_l,\Phi_l).
\end{equation} 
By the strong law of large numbers the individual matrix elements of this estimator converge 
to the matrix elements of the true parameter $\rho$ and has the advantage that it can be 
computed in real time. 
The disadvantages are that the matrix $\hat{\rho}^{(n)}$ as a whole need not 
be positive, normalized or even trace-class, and one has no control on the convergence 
$\hat{\rho}^{(n)}\to\rho$ in any relevant distance such as for example $\|\cdot\|_1$ 
due to the infinite number of matrix elements and ranges of the pattern functions 
$f_{i,i+d}$ increasing with $i$ and $d$. We can avoid this problem by choosing 
$\hat{\rho}^{(n)}$ to be an effectively finite dimensional selfadjoint matrix of dimension $N(n)$ 
growing with $n$ that is, $\hat{\rho}^{(n)}_{i,i+d}=0$ for $i+d> N(n)$, and 
$\hat{\rho}^{(n)}_{i,i+d}$ given by (\ref{eq.unbiased_estimator}) for $i+d\leq N(n)$. 
We apply now Hoeffding's inequality for the matrix elements, 
\begin{equation}\label{eq.hoeffding.matrix.elements}
\mathbb{P}(|\hat{\rho}^{(n)}_{i,i+d}-\rho_{i,i+d}|\geq a )\leq 
\mathrm{exp}\left(\frac{-na^2}{\|f_{i,i+d}\|_\infty^2}\right),
\end{equation} 
and let $\rho^{(n)}$ denote the restriction of the true density matrix to $N(n)$ dimensional 
subspace on which $\hat{\rho}^{(n)}$ is non-trivial. 
We will look at the $\|\cdot\|_2$-distance defined in general by 
\begin{equation*}
\|\tau-\tau'\|_2^2:=\mathrm{Tr}(|\tau-\tau'|^2)=\sum_{j,k\geq 0}|\tau_{j,k}-\tau^\prime_{j,k}|^2.
\end{equation*} 
Then from (\ref{eq.hoeffding.matrix.elements}) we obtain
\begin{equation}\label{eq.sum.probabilities}
\mathbb{P}(\|\hat{\rho}^{(n)}-\rho^{(n)}\|_2 \geq a )\leq 
N(n)^2\mathrm{exp}\left(\frac{-na^2}{\sum_{k,j=0}^{N(n)}\|f_{k,j}\|_\infty^2}\right).
\end{equation}
\begin{lemma}
The following holds:
\begin{equation}\label{eq.sum.pattern.functions}
\sum_{k\geq j\geq0}^N\|f_{k,j}\|_\infty^2 =O(N^{7/3}).
\end{equation}
\end{lemma}

\noindent{\it Proof.} We refer to the paper \cite{Leonhardt.Richter} for a more detailed 
analysis of the functions $\psi_k,\varphi_j$ and we mention here only some 
qualitative features. Let $\epsilon >0$ be fixed. The 
Plancherel-Rotarch formulas \cite{Szego} give asymptotic formulas for 
$\psi_k$ and $\varphi_k$ in three regions of $\R$: the ``classical region'' 
$|x|\leq \epsilon \sqrt{2k+1}$ where both have an oscillatory behavior and have absolute values 
bounded by the envelope function $\sqrt{2/\pi}(1-x^2)^{-1/4}$, the 
``classically forbidden region'' $|x|\geq \epsilon \sqrt{2k+1}$ in which $\psi_k$ decays as 
$x^ke^{-x^2/2}$ while $\varphi_k$ grows as $x^{-k-1}e^{x^2/2}$, and the ``transition region'' 
with width $\epsilon k^{-1/6}$ centered around the turning point $\sqrt{2k+1}$ in which 
\begin{equation} 
\psi_k(x)=2^{1/4} k^{-1/12} 
\mathrm{Ai}\left(\sqrt{2}k^{1/6}(x-\sqrt{2k+1})\right)
\end{equation}
and similarly for $\varphi_k$ with the Airy function \cite{Abramowitz} $\mathrm{Ai}$ replaced by 
$\mathrm{Bi}$.

The range of the pattern functions $f_{k,j}$ increases slowly with the distance to the diagonal 
$j-k$, thus the main contribution in (\ref{eq.sum.pattern.functions}) is brought by terms which 
lie 
away from the diagonal. Let $C$ be a fixed 
constant, then for the pattern function $f_{k,j}$ situated in the upper corner 
$j\geq C k$, the maximum is attained in the overlap of the classical region for $\varphi_j$ 
with the transition region of $\psi_k$, and can be estimated by using the Plancherel-Rotarch 
formulas  
\begin{equation}
\|f_{k,j}\|_\infty=O\left(\frac{j^{1/4}}{k^{1/12}}\right).
\end{equation}
We sum now over the upper corner to obtain asymptotic behavior of the sum 
(\ref{eq.sum.pattern.functions}).

\qed 

\vspace{4mm}

\noindent
In particular we have the following necessary 
condition for the $\|\cdot\|^2$-consistency:
\begin{equation}\label{eq.rate.N.unbiased.estimator}
n^{-1}N(n)^{7/3}\to 0, \qquad \mathrm{as}\qquad n\to \infty.
\end{equation}

\begin{theorem} Let $(\epsilon_n,N(n))$ be such that $\epsilon_n\to 0$, $N(n)\to\infty$ and 
\begin{equation}
\frac{n\epsilon_n^2}{N(n)^{7/3}}- 2\log N(n)\to\infty. 
\end{equation}
Then
\begin{equation}
\|\hat{\rho}^{(n)}-\rho\|_2^2=\|\rho^{(n)}-\rho\|_2^2+O_\mathbb{P}(\epsilon_n^2).
\end{equation}
Moreover if
\begin{equation}
\sum_{n=1}^\infty \mathrm{exp}\left(-\frac{n\epsilon_n^2}{N(n)^{7/3}}+2\log N(n)\right)<\infty
\end{equation}
then $\|\hat{\rho}^{(n)}-\rho\|_2\to 0$ almost surely.
\end{theorem}

\noindent
\textit{Proof.} The first statement follows directly from (\ref{eq.sum.probabilities}) and 
the fact that 
$\|\hat{\rho}^{(n)}-\rho\|_2^2=\|\rho^{(n)}-\rho\|_2^2+ \|\hat{\rho}^{(n)}-\rho^{(n)}\|_2^2$. 
The almost sure convergence follows from the first Borel-Cantelli lemma.

\qed

\vspace{4mm}

The homodyne tomography as presented in the beginning of this section does not take into account 
various losses (mode mismatching, failure of detectors) in the detection process which modify the 
distribution of results in a real measurement compared with the idealized case.  
Fortunately, an analysis of such losses \cite{Leonhardt} shows that they can be quantified 
by a single {\it efficiency} coefficient $0<\eta<1$ and the change in probability 
distributions amounts replacing $X_i$ by 
\begin{equation}
X_i':=\sqrt{\eta}X_i+\sqrt{(1-\eta)/2}Y_i
\end{equation}  
with $Y_i$ a sequence of i.i.d. standard Gaussian independent of all $X_j$. 
The efficiency-corrected probability density is then
\begin{equation}
p_\rho(y,\phi;\eta)=(\pi(1-\eta))^{-1/2}
\int_{-\infty}^\infty p(x,\phi)\mathrm{exp}\left[-\frac{\eta}{1-\eta}
(x-\eta^{-1/2}y)^2\right] ~dx.
\end{equation}
The problem is again the inference of the parameter $\rho$ from $(X_1',\Phi_1),(X_2',\Phi_2)$. 
One could follow two routes: use a deconvolution technique for the variable $X$ to obtain 
$p_\rho$ and then apply the previous kernel estimator for $\rho$, or find new pattern functions 
$f_{k,j}(x;\eta)$ such that
\begin{equation}\label{eq.quantum_tomographic_rho_n,n+d_eta}
\rho_{k,j}=\int_{-\infty}^\infty dx \int_0^\pi \frac{d\phi}{\pi} p_\rho(x,\phi;\eta)
f_{k,j}(x;\eta). 
\end{equation}    
Such functions are analyzed in \cite{D'Ariano.1,D'Ariano.2} where it is argued that 
the the method has a fundamental limitation for $\eta\leq 1/2$ in which case the 
pattern functions are unbounded, while for $\eta>1/2$ numerical calculations show 
that their range grows exponentially fast with both indices $j,k$. However there exists no 
proof of the conjecture which is implicitly made in the literature that it is impossible to 
estimate $\rho$  consistently for $\eta\leq 1/2$. A third route is to first estimate an 
intermediary state $\rho^{(\textrm{meas})}$ as in the $\eta=1$ case, and then to obtain 
$\rho$ from $\rho^{(\textrm{meas})}$ by inverting a {\it Bernoulli transformation} 
\cite{Leonhardt}:
\begin{equation}
\begin{CD}
p_\rho(\cdot,\cdot;\eta)@>{f_{k,j}}>> \rho^{(\textrm{meas})}
@>{\mathrm{inverse~Bernoulli}}>>\rho.\\ 
\end{CD}
\end{equation}    
To understand the (inverse) Bernoulli transformation let us consider first the diagonal elements 
$\{p_k=\rho_{k,k},~k=0,1..\}$ and 
$\{q_j=\rho^{(\textrm{meas})}_{j,j},~j=0,1..\}$ which are 
both probability distributions over $\N$ and represent the statistics of the number of 
photon in the two states. Let $ b_k^{k+p}=\binom{k+p}{k}\eta^k(1-\eta)^p$ be the 
binomial distribution. Then
\begin{equation}
q_j=\sum_{k=j}^\infty b_j^{k}(\eta)p_k
\end{equation}  
which is interpreted as result of the ``absorption'' process by which each photon is allowed 
to pass with probability $\eta$ and absorbed with probability $1-\eta$. The general formula is 
\begin{equation}
\rho^{(\textrm{meas})}_{j,k}=
\sum_{p=0}^{\infty}\left[ b_j^{j+p}(\eta) b_k^{k+p}(\eta)\right]^{1/2}\rho_{j+p,k+p},
\end{equation}  
and its inverse is obtained by replacing $\eta$ with $\eta^{-1}$! 
For $\eta\leq 1/2$ the power series $(1-\eta^{-1})^k$ appearing in the inverse transformation 
diverges, reflecting the obstruction for obtaining bounded pattern functions  $f_{k,j}(x;\eta)$.

\section{Sieve maximum likelihood estimation}
\label{sec.mle}

In this section we will develop a maximum likelihood 
approach to the estimation of the state $\rho$. Let us remind the reader of 
the terms of the problem: 
we are given a sequence $(X_1,\Phi_1),(X_2,\Phi_2)\dots$ of i.i.d. random variables with values in 
$\R\times[0,\pi]$ with probability density $p_\rho$ with respect to the Lebesgue measure 
$dx\times\frac{d\phi}{\pi}$ depending on the parameter $\rho\in\mathcal{S}(\mathcal{H})$. 
When taking into consideration the efficiency $\eta<1$ we have replace $p_\rho$ by 
$p_\rho(\cdot,\cdot;\eta)$. We would like to find 
\begin{equation*}
\hat{\rho}^{(n)}=\hat{\rho}^{(n)}(X_1,\Phi_1,\dots, X_n,\Phi_n),
\end{equation*} 
such that the $\|\cdot\|_1$-consistency holds:  
\begin{equation*}
\lim_{n\to\infty} \|\hat{\rho}^{(n)}-\rho\|_1=0, \qquad \text{a.s.}.
\end{equation*} 
Let $\hat{p}_n:=p_{\hat{\rho}^{(n)}}$ be the corresponding probability density. 
We denote by
\begin{equation}\label{eq.hellinger.distance}
h(P_1,P_2):=\left(\int(\sqrt{p_1}-\sqrt{p_2})^2~d\mu\right)^{1/2},
\end{equation}
the Hellinger distance 
between two probability distributions on $(\Omega,\Sigma,\mu)$ with densities $p_1,p_2$ 
with respect to $\mu$. The following relations are well known
\begin{equation}
\frac{1}{2}d_{\text{tv}}(P_1,P_2)\leq h(P_1,P_2)\leq \sqrt{d_{\text{tv}}(P_1,P_2)},
\end{equation}
and combined with (\ref{eq.contactivity.measurements}) give in the case of our measurement 
\begin{equation}\label{contractivity.T}
h(P_{\tau}, P_{\tau'})\leq \sqrt{\|\tau-\tau'\|_1}
\end{equation}
for arbitrary states $\tau,\tau'\in\mathcal{S}(\mathcal{H})$. As a consequence, the 
Hellinger consistency 
\begin{equation}
\lim_{n\to\infty} h(\hat{P}_n,P_\rho)\to 0, \qquad \text{a.s.},
\end{equation}
is weaker than the $\|\cdot\|_1$-consistency.

\vspace{4mm}

The maximum likelihood estimator is usually defined as the parameter $\tau$ which maximizes the 
log-likelihood $\sum_{a=1}^n \log p_\tau(X_a,\Phi_a)$. In this case the maximum is not 
achieved over the whole space and it seems more appropriate to restrict the attention to a 
subspace $\mathcal{Q}(n)$ on which the maximum exists, whose size grows with the number of data 
and such that $\cup_{n\geq 1}\mathcal{Q}(n)$ is dense in $\mathcal{S}(\mathcal{H})$ 
in the norm topology. Such a method is called sieved maximum likelihood and we refer to 
\cite{Geer, Wong&Shen} for the general theory. The choice of the spaces $\mathcal{Q}(n)$ should be 
tailored according to the problem one wants to solve,  the class of states one is interested in, 
etc. We will use here the {\it number states sieves} for which $\mathcal{Q}(n)$ consists of 
density matrices over the subspace spanned by the basis vectors 
$\psi_0,\dots,\psi_{N(n)}$ defined in (\ref{eq.psi_n}), 
with $N(n)$ an increasing function of $n$ which will be fixed later:  
\begin{equation}
\mathcal{Q}(n)=\left\{\tau\in\mathcal{T}_1(\mathcal{H})~:~ \tau_{j,k}=0 ~\text{for all}~ 
j>N(n) ~\text{or}~ k>N(n)\right\}. 
\end{equation}
The dimension of the space $\mathcal{Q}(n)$ is $N(n)^2$. Let 
\begin{equation}
\hat{\rho}^{(n)}:=\text{arg}\max_{\tau\in\mathcal{Q}(n)}\sum_{a=1}^n\log p_\tau(X_a,\Phi_a),
\end{equation}
and notice that by compactness arguments the maximum always exists. 

We define the convex map 
\begin{equation}
\mathbf{T}:
\mathcal{S}(\mathcal{H})\ni \tau\longmapsto p_\tau\in 
L_1(\R\times [0,\pi], dx\times \frac{d\phi}{\pi}),
\end{equation}
whose image $\mathcal{P}$ is the class of probability densities of the form 
(\ref{eq.concrete.p}), but for the moment we lack a more intrinsic characterization of 
its elements. 
The image of the sieve $\mathcal{Q}(n)$ is the convex hull $\mathcal{P}(n)$ of 
densities of the form 
\begin{equation}
p_\psi(x,\phi)=\left|\sum_{k=0}^{N(n)}\alpha_ke^{ik\phi}\psi_k(x)\right|^2,
\end{equation}
with $\psi=\sum_{k=0}^{N(n)}\alpha_k\psi_k$ a unit vector.

In order to obtain results on consistency of estimators, it is essential to bound the 
``size'' of the sieve by entropy numbers which we define here for with respect to 
the $\|\cdot\|_1$-distance. 

\begin{definition}
Let $\mathcal{F}$ be a class of probability densities. 
Let $N_{B,1}(\delta,\mathcal{F})$ be the smallest value of $p\in\N$ for which there 
exist pairs of functions $\{[f_j^L,f_j^U]\}$ with $j=1,\dots,p$ such that 
$\|f_j^L-f_j^U\|_1\leq\delta$ for all $j$, and such that for each 
$f\in\mathcal{F}$ there is a $j=j(f)\in\{1,\dots, p\}$ such that
\begin{equation*}
f_j^L\leq f \leq f_j^U.
\end{equation*}
Then $H_{B,1}(\delta, \mathcal{F})=\log N_{B,1}(\delta, \mathcal{F})$ is called 
{\it $\delta$-entropy with bracketing} of $\mathcal{F}$.
\end{definition} 
We note that this definition relies on the concept of positivity and distance between 
$L_1$-functions. 
But the same notions exist for the space of trace-class operators $\mathcal{T}_1(\mathcal{H})$, 
thus by replacing probability densities with 
density matrices and functions with selfadjoint trace class operators we obtain the definition of 
the $\delta$-entropy with bracketing $H_{B,1}(\delta, \mathcal{Q})$ 
for some space of density matrices $\mathcal{Q}$.
\begin{proposition}
Let $\mathcal{Q}(n)$ be the class of density matrices of dimension $N(n)$. Then
\begin{equation}\label{eq.bound.bracheting.entropy.norm1}
H_{B,1}(\delta,\mathcal{Q}(n))\leq C N(n)^2\log\frac{N(n)}{\delta}. 
\end{equation}
for some constant $C$ independent of $n$ and $\delta$.
\end{proposition}

\noindent \textit{Proof.}

Let $\{\rho_j, ~j=1,\dots, c(\delta,n)\}$ be a maximal set of density matrices in $\mathcal{Q}(n)$ 
such that for any $j\neq k$ we have $\|\rho_j-\rho_k\|_1 > \frac{\delta}{2N(n)}$. 
We define
\begin{equation*}
\rho_j^U=\rho_j + \frac{\delta}{2N(n)}\mathbf{1},  \qquad \qquad
\rho_j^L=\rho_j - \frac{\delta}{2N(n)}\mathbf{1}.
\end{equation*}
Then for any $\rho$ in the ball $B_1(\rho_j,\frac{\delta}{2N(n)})$ we have 
$\rho-\rho_i\leq  \frac{\delta}{2N(n)}\mathbf{1}$, thus
\begin{equation*}
\rho_j^L \leq \rho\leq \rho_j^U,
\end{equation*}
and clearly $\|\rho_j^L - \rho_j^U\|_1=\delta.$ 
It remains to estimate the number of balls $c(\delta,n)$. 
From standard arguments on dimension we obtain
\begin{equation*}
c(\delta,n)(\frac{\delta}{4N(n)})^{N(n)^2}\leq (1+\frac{\delta}{4N(n)})^{N(n)^2}- 
(1-\frac{\delta}{4N(n)})^{N(n)^2}
\end{equation*}
where the difference on the right side represents the volume between two balls of radii 
$1-\frac{\delta}{4N(n)}$ and $1+\frac{\delta}{4N(n)}$. As a rough estimation we obtain 
\begin{equation*}
c(\delta,n)\leq (1+\frac{4N(n)}{\delta})^{N(n)^2} \leq \left(5\frac{N(n)}{\delta}\right)^{N(n)^2},
\end{equation*}
The bracketing entropy is at most $\log c(\delta,n)$ and we obtain 
(\ref{eq.bound.bracheting.entropy.norm1}) with $C=1+\log 5$.

\qed

\begin{corollary}
Let $\mathcal{P}(n)^{1/2}$ be the class of $L_2$-functions 
$\{\sqrt{p_\rho}~:~p_\rho\in\mathcal{P}(n)\}$ and $H_{B}(\delta,\mathcal{P}(n)^{1/2})$ be the 
bracketing entropy with the $\|\cdot\|_2$-distance. Then
\begin{eqnarray}
&&H_{B,1}(\delta,\mathcal{P}(n))\leq N(n)^2\log\frac{N(n)}{\delta}\\
&&H_{B}(\delta,\mathcal{P}(n)^{1/2})\leq N(n)^2\log\frac{N(n)}{2\delta^2}
\label{bracketing.entropy.Hellinger}
\end{eqnarray}
\end{corollary}

\noindent{\it Proof.}

Let $\tilde{\mathbf{T}}$ be the linear extension to $\mathcal{T}_1(\mathcal{H}) $ 
of the map $\mathbf{T}$. Then $\tilde{\mathbf{T}}$ is positivity preserving that is, for any 
$\tau,\tau'\in \mathcal{T}_1(\mathcal{H})$ such that $\tau\geq \tau'$ then 
$p_\tau\geq p_{\tau'}$ where we extend the notation $p_\tau=\tilde{\mathbf{T}}(\tau)$ to all 
trace-class operators. Let $[\rho_j^U, \rho_j^L]$ be the $\delta$-bracketing matrices from the 
previous proposition. Then by the above observation 
$[\tilde{\mathbf{T}}(\rho_j^U),\tilde{\mathbf{T}}(\rho_j^U)]$ is a set of brackets for 
$\mathcal{P}(n)=\mathbf{T}(\mathcal{Q}(n))$. From the monotonicity on the $\|\cdot\|_1$ proved 
(\ref{eq.contactivity.measurements}) we obtain 
$\|\tilde{\mathbf{T}}(\rho_j^U)-\tilde{\mathbf{T}}(\rho_j^U)\|_1\leq \delta$.

For the second inequality we note that 
$[\tilde{\mathbf{T}}(\rho_j^U)^{1/2},(\tilde{\mathbf{T}}(\rho_j^U)_+)^{1/2}]$ is a set of 
brackets for $\mathcal{P}(n)^{1/2}$ and then it can be shown than 
\begin{equation*}
\|\tilde{\mathbf{T}}(\rho_j^U)^{1/2}-(\tilde{\mathbf{T}}(\rho_j^U)_+)^{1/2}\|_2^2\leq \frac{\delta}{2}.
\end{equation*}

\qed

We will concentrate now on the Hellinger consistency of the sieve maximum likelihood 
estimator $\hat{P}_n$. We will appeal to a theorem from \cite{Wong&Shen}, which is similar to 
other results in the literature on non-parametric $M$-estimation (see for example \cite{Geer}). 
There are two competing factors which contribute to the connvergece of $h(\hat{P}_n,P_\rho)$. 
The first is related with the approximation properties of the sieves with respect to the whole 
parameter space. Such a ``distance'' from $\rho$ to the sieve $\mathcal{Q}(n)$ can take different 
expressions, for example in terms of Kullback-Leibler divergence 
$K(q,p):=\int p\log\frac{p}{q}$,    
\begin{equation}
\delta_n(0+):=\underset{\rho'\in\mathcal{Q}(n)}{\mathrm{inf}}K(p_{\rho'},p_\rho). 
\end{equation}  
and 
\begin{equation}
\tau_n=\lim_{k\to\infty}\int p_\rho(\log \frac{p_\rho}{p_k})^2
\end{equation}  
where $\{p_k, k=1,2,\dots\}\subset \mathcal{P}(n)$ is a sequence such that 
$\lim_{k\to\infty}K(q_k,p_\rho)= \delta_n(0+)$. Another natural rate which will be used later 
is  
\begin{equation}
\gamma_n=\underset{\rho'\in\mathcal{Q}(n)}{\mathrm{inf}}\|\rho-\rho'\|_1.
\end{equation}
Notice that all this numbers depend on the growth rate of the sieve $N(n)$. 
The second factor influencing the convergence of $h(\hat{P}_n,P_\rho)$ is the size of the sieves 
which is expresses by the bracketing entropy. The non-parametric m.l. estimation theory shows that 
the following {\it entropy integral inequality} plays an important role in determining the rate of 
convergence
\begin{equation}\label{eq.entropy.integral.inequality}
J_B(\delta, \mathcal{P}^{1/2}(n)):=\int_{\delta^2/2^8}^{\sqrt{2}\delta} ~
H_B^{1/2}(\frac{u}{c_3}, \mathcal{P}(n)^{1/2})du\leq c_4\sqrt{n}\delta^2.
\end{equation}

\begin{theorem}\label{th.Wong&Shen.sieved}
There exist constants $c_i, i=1,\dots, 4$ such that if $\delta_n$ is the smallest 
value satisfying (\ref{eq.entropy.integral.inequality}) and we define 
\begin{displaymath}
\epsilon_n=
\left\{ \begin{array}{ll}
\delta_n,    & \mathrm{if}~\delta_n(0+)<\frac{1}{4}c_1 \delta_n \\              
\left(4\delta_n(0+)/c_1\right)^{1/2},          & \mathrm{otherwise}         
\end{array} \right.
\end{displaymath}
then 
\begin{equation}\label{eq.probability_inequality}
\mathbf{P}\left( h(\hat{P}_n,P_\rho)\geq \epsilon_n\right)\leq 
5 e^{-c_2n\epsilon_n^2}+ \frac{4\tau_n}{c_1 n \epsilon_n^2}.
\end{equation}
\end{theorem}

In calculating the entropy integral we take into account (\ref{bracketing.entropy.Hellinger}),
\begin{eqnarray}\label{eq.entropy.integral}
J_{B}(\delta, \mathcal{P}^{1/2}(n)) &=&
O\left[N(n)\int_{\delta^2}^\delta ~\left(\log \frac{N(n)^{1/2}}{u}\right)^{1/2}du\right]
\nonumber\\
&=&O\left[N(n)^{3/2}\int_{N(n)^{1/2}/\delta}^{N(n)/\delta^2} 
w^{-2}(\log w)^{1/2}dw\right]\nonumber\\
&=&  
O\left[N(n)\delta\left(\log \frac{N(n)}{\delta}\right)^{1/2}\right].
\end{eqnarray}
From the entropy inequality we obtain the rate $\delta_n$ satisfying 
\begin{equation}\label{eq.rate.delta_n}
\frac{N(n)}{\delta_n}=O\left(\sqrt{\frac{n}{\log n}}\right).
\end{equation}

\begin{theorem}\label{th.Hellinger.consistency}
Suppose that the state $\rho$ satisfies $\tau_n=O(N(n)^{-\tau})$ for some $\tau>0$. 
Let $\hat{\rho}^{(n)}$ be the sieve MLE with $N(n)=o((\frac{n}{\log n})^{1/2})$ and 
$N(n)^{-1}=o(n^{-\theta})$ for some $\theta>0$. Then $\hat{p}_n$ is Hellinger consistent, i.e.
\begin{equation}
h(\hat{P}_n,P_\rho)\to 0\qquad \mathrm{a.s.}. 
\end{equation}
\end{theorem}

\noindent\textit{Proof.} We apply Theorem \ref{th.Wong&Shen.sieved} to our particular situation. 
We can choose a rate $\delta_n\to 0$ satisfying (\ref{eq.rate.delta_n}) for our particular 
choice of $N(n)$ and decreasing slower that $1/\log n$. Then 
\begin{equation*}
\sum_{n=1}^\infty\left( 
5e^{-c_2 n\epsilon_n^2}+\frac{4\tau_{n}}{c_1 n(\epsilon_n)^2} 
\right)<\infty
\end{equation*} 
because the lower bound for $N(n)$ and the class assumption imply that $\tau_n$ decreases 
faster than some power of $n$. A standard application of the first Borel-Cantelli lemma proves 
almost sure convergence of $h(\hat{P}_n,P)\to 0$.

\qed

From the physical point of view, we are more interested in the convergence of the state estimator 
$\hat{\rho}^{(n)}$ which is in principle a stronger requirement than Hellinger consistency. 
We will show however that the two are equivalent  by applying a quantum analogue of the 
classical Scheff\'{e}'s lemma \cite{Williams} 
which says that if a sequence of probability densities converge pointwise almost everywhere 
to a probability density, then they also converge in $\|\cdot\|_1$. We will replace the 
$L_1$ space by the space of trace-class operators $\mathcal{T}_1(\mathcal{H})$, and the 
pointwise convergence by {\it weak operator convergence} which is roughly 
$\br\psi,\mathbf{X}_n\psi\ke\to\br\psi,\mathbf{X}\psi\ke$ for all $\psi\in\mathcal{H}$. 
In particular for density matrices it is sufficient to check the individual 
convergence of all matrix elements in a given basis. For the proof and other non-commutative 
convergence theorems we refer to \cite{Simon.3}.  

\begin{theorem}\label{cor.scheffe}
Let $\rho_n$ be a sequence of density matrices converging weakly to another density 
matrix $\rho$. Then $\|\rho_n-\rho\|_1\to 0$ as $n\to\infty$.
\end{theorem}

\begin{corollary}
The Hellinger consistency of $\hat{P}_n$ is equivalent to the $\|\cdot\|_1$-consistency of 
$\hat{\rho}^{(n)}$. In particular, under the assumptions of Theorem \ref{th.Hellinger.consistency} 
we have $\|\hat{\rho}^{(n)}-\rho\|_1\to 0$, a.s..  
\end{corollary}

\noindent{\it Proof.} By Theorem \ref{cor.scheffe} it is enough to prove almost sure convergence 
of each matrix element individually. But we have shown in (\ref{eq.quantum_tomographic_rho_n,n+d}) 
that $\rho_{k,j}$ and $\hat{\rho}^{(n)}_{k,j}$ can be expressed as the integral of $ p_\rho$ and 
respectively $\hat{p}_n$ with bounded pattern functions $f_{k,j}(x)e^{-i(j-k)\phi}$. 

\qed

\vspace{4mm}

\noindent{\bf Concluding Remarks.} 
There are many open questions related to quantum tomography and we would like to enumerate a few 
of them here. 

The equivalence in last corollary holds as well for efficiency 
$\eta>\frac{1}{2}$ as we only use the fact that the pattern functions are bounded, but seems to 
fail for $\eta\leq\frac{1}{2}$ when the pattern functions are unbounded. 
Is $\eta=\frac{1}{2}$ some kind of transition point between two 
convergence regimes? 

Another problem which has not been treated here is that of rates of 
convergence for estimators. A possible way to obtain this is to find the rates $\epsilon_n$ 
of convergence for $h(\hat{P}_n,P_\rho)$ and then to use the modulus of continuity 
$\omega_n(\epsilon)$ of the inverse map on the sieves
\begin{equation}
\mathbf{T}^{-1}:\mathcal{P}(n)\to \mathcal{Q}(n)
\end{equation}   
to obtain the rough rate $\omega_n(\epsilon_n)$ for $\|\hat{\rho}^{(n)}-\rho\|_1$. 
This will lead to a slower increase of the sieve dimension $N(n)$. Is there a more direct 
approach to the estimation of the rates? Does the maximum likelihood estimator converge faster 
than the kernel estimator using pattern functions presented in section \ref{sec.qhomodyne}? 
Can we use penalization instead of arbitrarily choosing the dimension of the sieve?  

On the practical side of the problem, finding the maximum of the likelihood function over 
a set of density matrix is non-trivial. The positivity and normalization constraints must be 
taken into account. 

In the case $\eta<1$ we have to deconvolve the noise introduced by the detection imperfection. 
The analysis made for perfect detection should be made also in this case. It seems to us that 
the conjecture made by D'Ariano referring to the impossibility of reconstructing the state for 
$\eta\leq \frac{1}{2}$ is not true in general, but it does pose a kind of restriction. 
One should identify the class of states for which the reconstruction is still possible.

Needless to say, the methods used here for quantum tomography can be applied in other problems 
of quantum estimation, such as for example estimating how certain devices transform the states of 
quantum systems.

\bibliography{tomobib}

\begin{thebibliography}{10}

\bibitem{Abramowitz}
{Abramowitz, M.}, {Stegun, I.A.}, {\em Handbook of Mathematical Functions\/},
  National Bureau of Standards, Washington (1972).

\bibitem{Barndorff-Nielsen&Gill&Jupp}
{Barndorff-Nielsen, O.E.}, {Gill, R.}, {Jupp, P.E.}, On quantum statistical
  inference, to appear in J. Royal Stat. Soc. B.

\bibitem{D'Ariano.1}
D'Ariano G., Tomographic measurement of the density matrix of the radiation
  field, {\em Quantum Semiclass. Optics\/}, {\bf 7}, (1995), 693--704.

\bibitem{D'Ariano.10}
D'Ariano G., Quantum tomography: general theory and new experiments, {\em
  Fortschr. Phys.\/}, {\bf 48}, (2000), 579--588.

\bibitem{D'Ariano.5}
D'Ariano G., Tomographic methods for universal estimation in quantum optics, in
  {\em International School of Physics Enrico Fermi\/}, volume 148, IOS Press
  (2002).

\bibitem{D'Ariano.2}
{D'Ariano, G.M.}, {Leonhardt, U.}, {Paul, H.}, Homodyne detection of the
  density matrix of the radiation field, {\em Phys. Rev. A\/}, {\bf 52},
  (1995), R1801--R1804.

\bibitem{D'Ariano.0}
{d'Ariano, G.M.}, {Macchiavello, C.}, {Paris, M.G.A.}, Detection of the density
  matrix through optical homodyne tomography without filtered back projection,
  {\em Phys. Rev. A\/}, {\bf 50}, (1994), 4298--4302.

\bibitem{D'Ariano.8}
{D'Ariano, G.M.}, {Maccone, L.}, {Paris, M.G.A.}, Quorum of observables for
  universal quantum estimation, {\em J. Phys. A\/}, {\bf 35}, (2001), 93--103.

\bibitem{Holevo}
Holevo A., {\em Probabilistic and Statistical Aspects of Quantum Theory\/},
  North-Holland (1982).

\bibitem{Leonhardt}
Leonhardt U., {\em Measuring the Quantum State of Light\/}, Cambridge
  University Press (1997).

\bibitem{Leonhardt.Richter}
{Leonhardt, U.}, {Munroe, M.}, {Kiss, T.}, {Richter, Th.}, {Raymer, M.G.},
  Sampling of photon statistics and density matrix using homodyne detection,
  {\em Optics Communications\/}, {\bf 127}, (1996), 144--160.

\bibitem{D'Ariano.3}
{Leonhardt, U.}, {Paul, H.}, {D'Ariano, G.M.}, Tomographic reconstruction of
  the density matrix via pattern functions, {\em Phys. Rev. A\/}, {\bf 52},
  (1995), 4899--4907.

\bibitem{Simon.3}
Simon B., {\em Trace Ideals and their Applications\/}, Cambridge University
  Press (1979).

\bibitem{Smithey}
{Smithey, D.T.}, {Beck, M.}, {Raymer, M.G.}, {Faridani, A.}, Measurement of the
  Wigner distribution and the density matrix of a light mode using optical
  homodyne tomography: Application to squeezed states and the vacuum, {\em
  Phys. Rev. Lett.\/}, {\bf 70}, (1993), 1244--1247.

\bibitem{Szego}
Szeg\"{o} G., {\em Orthogonal Polynomials\/}, Americam Mathematical Society,
  Providence (1975).

\bibitem{Geer}
van~de Geer S., {\em Applications of Empirical Process Theory\/}, Cambridge
  University Press (2000).

\bibitem{Williams}
Williams D., {\em Probability with Martingales\/}, Cambridge University Press
  (1991).

\bibitem{Wong&Shen}
{Wong, W.H.}, {Shen, X.}, Probability inequalities for likelihood rations and
  convergence rates of sieve MLEs, {\em Ann. Statist.\/}, {\bf 23}, no.~2,
  (1995), 339--362.

\end{thebibliography}

\end{document}